\documentclass[onecolumn, notitlepage]{article}

\usepackage[utf8]{inputenc}
\usepackage{amsmath}
\usepackage{graphicx}
\usepackage{amsfonts}
\usepackage{tensor} 
\usepackage{enumerate}
\usepackage{tensor} % Nice tensors

\usepackage{empheq} % Package for emphasizing equations (with boxes)
\newcommand*\widefbox[1]{\fbox{\hspace{0em}#1\hspace{0em}}}

\usepackage{authblk} % Authors with different affiliations

\usepackage[backend=biber,style=numeric-comp,sorting=none]{biblatex}
\bibliography{a.bib}

\usepackage{caption}
\DeclareCaptionFormat{myformat}{#1#2#3\hrulefill}
\newcommand{\vect}[1]{\boldsymbol{\mathbf{#1}}} % Bold vectors for both latin and greek letters!
\usepackage{letltxmacro} % Package containing closed sqrts
\LetLtxMacro{\oldsqrt}{\sqrt}% makes all sqrts closed
\renewcommand{\sqrt}[1][\ ]{%
  \def\DHLindex{#1}\mathpalette\DHLhksqrt}
\def\DHLhksqrt#1#2{%
  \setbox0=\hbox{$#1\oldsqrt[\DHLindex]{#2\,}$}\dimen0=\ht0
  \advance\dimen0-0.2\ht0
  \setbox2=\hbox{\vrule height\ht0 depth -\dimen0}%
  {\box0\lower0.71pt\box2}}

\title{A source of the C-metric with perfect translational inertial dragging}
\date{}
\author[1]{Brynjar Arnfinnsson \thanks{brynjar.arnfinnsson@fys.uio.no}}
\author[1,2]{Øyvind Grøn \thanks{Oyvind.Gron@hioa.no}}
\affil[1]{Department of Physics, University of Oslo, 0316 Oslo, Norway}
\affil[2]{Oslo and Akershus University College of Applied Sciences, Faculty of Technology, Art and Design, P.O. Box 4 St. Olavs Plass, N-0130 Oslo, Norway}

\begin{document}
\maketitle
\begin{abstract}
A new source of the C-metric is described using Israel's formalism. This source is a singular accelerated shell. By construction, perfect inertial dragging is realized inside the shell. The equation of state and energy conditions for the shell are discussed.
\end{abstract}

\section{Introduction}
%The following article is a summary of the original work involved in the Ms.S-thesis \cite{ba}.\\ \\
Accelerating and rotating mass shells have been frequently used as models to study perfect inertial dragging - a condition necessary for the fulfilment of Mach's principle. This study is a continuation of this program. 
Excellent accounts for the history surrounding Mach's principle can be found in \cite{MP}, and a shorter version in \cite{onthehistory}. \\ \\
Inertial dragging (or ``frame dragging''') is the effect that inertial systems are dragged along in the direction of acceleration of a nearby accelerating mass. This means, that a free test particle or observer, subject only to the ``force'' (effect) of a nearby accelerated gravitating body, will, as viewed from a distant observer, accelerate in the same direction as the accelerated body. The accelerating test body or observer, will not feel anything. The testbody/observer is in free fall. If the freely falling observer carries an accelerometer, it will measure zero acceleration. This is because this observer is at rest relative to the spacetime surrounding him/her locally. Like a boat that flows with the same velocity as the river, the observer falls freely with the same acceleration as that of the ``river of space'' \cite{rivermodel}.\\ \\
In a preliminary gravitational theory Einstein in 1912 calculated that inside a massive accelerated shell the inertial frames would be dragged along in the same direction with magnitude $a_\text{inertial} = \frac{3M}{2R}a_\text{shell}$ \cite{onthehistory}. This was the first attempt at calculating a translational inertial dragging effect. Later the same effect has been considered by some authors \cite{davidson, groneriksen, menshikov, lyndenbell, lyndenbell2, pfister}.\\ \\
Pfister, Frauendiener and Hengge \cite{pfister} studied a charged shell accelerated by a dipolar charge distribution $\lambda\sigma(r)\sin\theta$. They studied first the weak field case (shell with small mass and charge), then an arbitrarily massive shell with a small charge, and lastly the general strong field case. In all cases the calculations were limited to first order in the dimensionless parameter $\lambda$. \\ \\In the case of a massive shell with a small charge, they found an explicit formula for the dragging coefficient, defined as:
\begin{equation}
d_{\text{linear}} = \frac{g}{b} \:,
\end{equation}
where $g$ and $b$ are the acceleration of a test particle inside the shell and the shell itself, respectively. See equations (48)-(49) and figure 3 in their paper. In the collapse limit they found that this factor goes to unity (with a horizontal tangent), which means that perfect inertial dragging is realized in this model (for arbitrary charge distributions $\sigma(r)$).\\ \\ In the general strong field case their set of differential equations is not separable, so they turned to a numerical procedure to solve the problem. They found that the dragging factor is exactly 1 only in the special case where $R = 2M$ (collapse limit) \emph{and} $q = 0$. Interpreting their contour plot of the dragging factor (figure 4 in their article) the dragging factor approaches unity as both $R\rightarrow 2M$ \emph{and} $q^2/R^2 \rightarrow 0$.\\ \\
The acceleration of the shell is $b$ and must be proportional to the product of the parameters $\lambda$ and $q$. This means that in all cases involved in this study, only first order terms in the acceleration $b$ is kept in the calculations. So in this sence these results are not exact, they only represent the limit where the acceleration of the shell is small.\\ \\They also analyzed the weak and dominant energy conditions for the shell. Perfect inertial dragging is only realized in the regions where the dominant energy condition is violated (compare their figures 4 and 1).\\ \\
We study a new model of an accelerated mass shell. For the first time we present a result that is analytic and exact to all orders of the acceleration parameter $\alpha$. This model casts light on the requirements for achieving perfect translational inertial dragging. In addition, it provides a source for the C-metric.\\ \\
In section 2 the basics of the Israel formalism is summarized. In section 3 we introduce the metric outside an accelerated black hole, which will serve as the exterior metric in our model. In section 4 the Israel formalism is applied to the spherical accelerated shell, and the shell is given a physical interpretation as a perfect fluid. In section 5 we discuss the properties of the shell further, and discuss the inertial dragging effect inside the shell. We conclude in section 6 and pose some remaining open questions that need further investigation.
\section{A short recap of Israel's formalism}
Following the formalism developed by W. Israel \cite{israel} for hypersurfaces in general relativity, we want to solve the Lanczos equation:
\begin{equation}
\boxed{ \epsilon \kappa S\indices{^a_{b}} = \left[ K\indices{^a_{b}} \right] - \delta\indices{^a_{b}} \left[ K \right]\:.}
\label{lanczos}
\end{equation}
Let the spacetime $\mathcal{M}$ be divided into two domains:
\begin{equation}
\mathcal{M} = \mathcal{M}^+ \cup \mathcal{M}^- \:,
\end{equation}
with a common hypersurface boundary: 
\begin{equation}
\Sigma = \partial \mathcal{M}^+  \cap \partial \mathcal{M}^- \:.
\end{equation}
$S\indices{^a_{b}}$ is the energy-momentum tensor of this hypersurface. Square brackets define the discontinuity operation:
\begin{equation}
[A] \equiv A^+ - A^-\:,
\end{equation}
where the $+$ means evaluated infinitesimaly outside $\Sigma$ in $\mathcal{M}^+$, etc. The extrinsic curvature can be calculated from
\begin{equation}
K^\pm_{\mu\nu} = \epsilon n_\alpha \Gamma\indices{^\alpha_{\mu\nu}} |^\pm \:,
\label{excurve}
\end{equation}
where $\Gamma\indices{^\alpha_{\mu\nu}}$ are the Christoffel symbols, and $n_\alpha$ is a normal vector to the hypersurface, with normalization:
\begin{equation}
\vect n \cdot \vect n = g_{\mu\nu} n^\mu n^\nu \equiv \epsilon = 
\begin{cases}
\hfill 1, & \text{if } \Sigma \text{ is timelike,} \\
\hfill -1, & \hfill \text{if } \Sigma \text{ is spacelike.}
\end{cases}
\end{equation}
A general tensor $A^{\alpha\beta}$ transforms to the coordinates intrinsic to the hypersurface by:
\begin{equation}
A_{ab} = A_{\alpha\beta}  e^\alpha_a e^\beta_b \:.
\label{intr_transf}
\end{equation}
Specifically, the induced metric tensor $h_{ab}$ is continuous, $\left[ h_{ab}\right] = 0$, and given by:
\begin{align}
h_{ab} = g_{\alpha\beta}^\pm e^{\alpha^\pm}_a e^{\beta^\pm}_b\:.
\label{continous_induced_metric}
\end{align}
The tangent vectors $e^\alpha_a$ in $\Sigma$ are given by:
\begin{equation}
e^\alpha_a = \frac{\partial x^\alpha}{\partial y^a} \:,
\end{equation}
when the parametrization of the hypersurface is given by:
\begin{equation}
x^\alpha = x^\alpha (y^a) \:,
\end{equation}
where $y^a$ are the intrinsic coordinates of the surface.

\section{An accelerated black hole}
\subsection{The C-metric}
A generalization of the Schwarzschild metric to an accelerated mass is known as the C-metric \cite{ehlers, kin_walker, bonnor, hong_teo, gkp, culetu}. The line-element is often given in the original form \cite{ehlers, kin_walker}:
\begin{equation}
ds^2 = \frac{1}{(\tilde{x}+\tilde{y})^2}\left( -\tilde{F} d\tilde{t}^2 + \frac{d\tilde{y}^2}{\tilde{F}} + \frac{d\tilde{x}^2}{\tilde{G}} + \tilde{G}d\tilde{z}^2 \right)\:,
\end{equation}
where $\tilde{F}(\tilde{y})$ and $\tilde{G}(\tilde{x})$ are cubic polynomials on the form
\begin{align}
\tilde{G}(\tilde{x}) = a_0 + a_1 \tilde{x} + a_2\tilde{x}^2 + a_3 \tilde{x}^3 \:,
\label{freedom}
\end{align}and
\begin{align}
\tilde{F}(\tilde{y}) = -\tilde{G}(-\tilde{y}) \:.
\end{align}
The choice of the constants $a_i$ are related to the choice of coordinates. To see this consider the coordinate transformation:
\begin{align}
\tilde{t} &= \frac{c_0}{A} \text{t} \:,\\
\tilde{y} &= Ac_0\text{y}-c_1\:, \\
\tilde{x} &= Ac_0\text{x} + c_1 \:,\\
\tilde{z} &=  \frac{c_0}{A} \phi \:.
\label{freedom_2}
\end{align}
The line-element will then be adjusted to:
\begin{align}
ds^2 = \frac{1}{A^2(\text{x}+\text{y})^2}\left( -\frac{\tilde{F} d\text{t}^2}{A^2} + \frac{A^2 d\text{y}^2}{\tilde{F}} + \frac{A^2 d\text{x}^2}{\tilde{G}} + \frac{\tilde{G}d\phi^2}{A^2}  \right)\:.
\end{align}
Let $\text{F} = A^{-2}\tilde{F}$ and $\text{G} = A^{-2}\tilde{G}$. This will cast the metric in another usual form\cite{kin_walker, bonnor}:
\begin{align}
ds^2 = \frac{1}{A^2(\text{x}+\text{y})^2}\left( -\text{F} d\text{t}^2 + \frac{d\text{y}^2}{\text{F}} + \frac{d\text{x}^2}{\text{G}} + \text{G}d\phi^2 \right)\:,
\end{align}
The coefficients of the functions $\text{G}$ and $\text{F}$ can now be adjusted by the choice of $c_1$ in \eqref{freedom_2} so that one of them is zero (except the cubic coefficient which does not depend on $c_1$). The standard choice has often been to choose $a_1 = 0$ and $a_0 = -a_2 = 1$ which gives the cubic polynomials the form
\begin{align}
\text{G} = 1 - \text{x}^2 - 2MA\text{x}^3 \:,\:\: \text{F} = -1 + \text{y}^2 - 2MA\text{y}^3 \:.
\end{align}
Hong and Teo \cite{hong_teo} used the freedom in \eqref{freedom} and \eqref{freedom_2} not to remove the linear terms, as has been the standard, but to make the root structure of the cubic polynomials as simple as possible. They arrive at the line-element:
\begin{align}
ds^2 = \frac{1}{\alpha^2(x+y)^2}\left(-F d\tau^2 + \frac{dy^2}{F} + \frac{dx^2}{G} + Gd\varphi^2\right) \:,
\end{align}
where
\begin{align}
G = (1-x^2)(1+2\alpha m x) \:, \:\: F = -(1-y^2)(1-2\alpha m y) \:.
\end{align}
Notice that the coordinates (t, y, x, $\phi$) have been rescaled to $(\tau, y, x, \varphi)$ (equations (10) and (11) in their paper), and the parameters $A$ and $M$ have been rescaled to $\alpha$ and $m$. In this form the coordinate $x$ is constrained to lie between -1 and 1, and we must have $0<2\alpha m < 1$ in order to preserve the signature of the metric.

\subsection{The C-metric in spherical coordinates}
We here summarize the relevant parts of the study of the C-metric by Griffiths, Krtou\v{s} and  Podolsk\'{y} \cite{gkp}. Make the coordinate transformation:
\begin{align}
x = \cos\theta \:,\:\:\: y = \frac{1}{\alpha\tau} \:, \:\:\: \tau = \alpha t \:.
\end{align}
The line-element becomes:
\begin{align}
ds^2 = & \frac{1}{(1+\alpha r \cos\theta)^2} \left(
-Q dt^2 +  \frac{dr^2}{Q} +  \frac{r^2 d\theta^2}{P} + Pr^2 \sin^2\theta d\varphi^2
\right) \:,
\end{align}
where the functions $Q$ and $P$ are given by:
\begin{align}
Q = &  \left( 1-\alpha^2 r^2 \right) \left( 1 - \frac{2m}{r} \right) \:, \\
P = & 1 + 2 \alpha m \cos\theta \:.
\end{align}
By inspection of $Q$, one sees that there are two coordinate singularties which occur at
\begin{align}
r &= 2m \:, \\
r &= \frac{1}{\alpha}\:,
\end{align}
where the first coordinate singularity corresponds to a black hole horizon, and the second corresponds to the horizon of a uniformly accelerating reference frame (``Rindler horizon''). In this way the metric can be viewed as a nonlinear combination of the Schwarzschild and Rindler spacetimes, thus representing the metric outside an accelerated point-particle or black hole. \\ \\In addition to the requirement $0<2\alpha m < 1$, we are only interested in the region of spacetime which lies inside the Rindler horizon:
\begin{equation}
r < \frac{1}{\alpha}\:.
\end{equation}
These are our constraints. Note that while the first constraint is a general one, the second one only cuts away a part of the spacetime which we will not deal with in this article.\\ \\The range of the $\varphi$ coordinate is $(-\pi C, \pi C)$. Griffiths \emph{et al.} consider the circumference to radius ratio for a small circle around the two half-axes $\theta = 0$ and $\theta = \pi$. In the first case the result is $2\pi C(1+2\alpha m)$ and in the second case it is $2\pi C(1-2\alpha m)$. Since these differ from $2\pi$ we have conical singularities along these half-axes (with different conicity). We see that choosing $C = (1\pm2\alpha m)^{-1}$ will remove one of these singularities, but not both at the same time. Griffiths \emph{et al.} choose $C = (1+2\alpha m)^{-1}$, removing the conical singularity at the $\theta = 0$ half-axis. They interpret the conical singularity at the $\theta = \pi$ half-axis as representing a \emph{``semi-infinite cosmic string under tension''}, and that the tension in the string is the cause of the force accelerating the Schwarzschild-like particle along the $\theta = \pi$ axis.\\ \\The range of the rotational coordinate can be rescaled to $2\pi$ by:
\begin{align}
\phi = C^{-1} \varphi \:.
\end{align}
With the above choice of the constant $C$ the line-element is now:
\begin{align}
ds^2 = & \frac{1}{(1+\alpha r \cos\theta)^2} \left(
-Q dt^2 +  \frac{dr^2}{Q} +  \frac{r^2 d\theta^2}{P} + \frac{Pr^2 \sin^2\theta}{(1+2\alpha m)^2} d\phi^2
\right) \:.
\end{align}
In order to simplify calculations later, we introduce the function $D$ in the metric tensor:
\begin{align}
D \equiv (1 + \alpha r \cos\theta)^2 \:.
\end{align}
The metric tensor can then be written:
\begin{align}
g_{\mu\nu} = & \frac{1}{D} \text{ diag} \left(
-Q, \frac{1}{Q}, \frac{r^2}{P}, \frac{Pr^2 \sin^2\theta}{(1+2\alpha m)^2}
\right) \:.
\label{c_metric}
\end{align}

%%%%%%%%%%%%%%%

\subsubsection{$\alpha \rightarrow 0$: The Schwarzschild limit}
Taking the limit where the parameter $\alpha$ goes to zero, the functions $Q$, $P$ and $D$ simplify:
\begin{align}
Q (\alpha = 0) = 1-\frac{2m}{r} \:,\: P (\alpha = 0) = 1\:,\: D(\alpha = 0) = 1 \:,
\end{align}
 and we recover the familiar Schwarzschild solution:
\begin{equation}
g_{\mu\nu} = \text{diag}\left( -\left(1-\frac{2m}{r}\right),\left(1-\frac{2m}{r}\right)^{-1}, r^2, r^2\sin^2\theta  \right)
\end{equation}
In this limit the parameter $m$ is the gravitational mass of the object. However we shall see in section 4 that this is not the case in the general scenario. This limit also serves as an important special case since our results are well known for the Schwarzschild spacetime.

\subsubsection{$m\rightarrow 0$: The weak field limit}
Taking the limit where the parameter $m$ goes to zero, the functions $Q$ and $P$ simplify as follows:
\begin{align}
Q (m = 0) = 1- \alpha^2 r^2 \quad,\quad P (m = 0) = 1\:,
\end{align}
and the metric reduces to:
\begin{align}
g_{\mu\nu}  & = \frac{1}{(1 + \alpha r \cos\theta)^2} \text{ diag} \left(
-(1- \alpha^2 r^2), \frac{1}{1- \alpha^2 r^2}, r^2, r^2 \sin^2\theta
\right) \:.
\end{align}
To clarify what type of spacetime this is, apply the transformation \cite{gkp}:
\begin{align}
\zeta = \frac{\sqrt{1-\alpha^2 r^2}}{\alpha(1+\alpha r \cos\theta} \:,\: \rho = \frac{r\sin\theta}{1+\alpha r \cos\theta} \:,\: \tau = \alpha t \:.
\label{rindler_tranfs}
\end{align}
Then the line-element reduces to:
\begin{align}
ds^2 = -\zeta^2 d\tau^2 + d\zeta^2 + d\rho^2 + \rho^2 d\phi^2 \:.
\end{align}
Which is the Rindler form of Minkowski spacetime in cylindrical coordinates. Applying the transformation
\begin{align}
T = \pm\zeta \sinh \tau \quad,\quad Z = \pm\zeta \cosh \tau \:,
\label{mink_transf}
\end{align}
one recovers the standard form of the Minkowski metric in cylindrical coordinates:
\begin{align}
ds^2 = -dT^2 + dZ^2 + d\rho^2 + \rho^2 d\phi^2 \:.
\end{align}
This shows that in the weak field limit the C-metric reduces to the metric of a uniformly accelerating reference frame. It follows from the knowledge of this special case that particles with a constant position in the coordinates of the C-metric ($r$, $\theta$ and $\phi$ constant) follow worldlines that in the Rindler coordinates are given by
\begin{equation}
Z^2 - T^2 = \frac{1-\alpha^2 r^2}{\alpha^2 (1+\alpha r \cos \theta)^2} \:.
\end{equation}
Particularly, the origin particle of the accelerated frame has the acceleration $\alpha$ in the positive or negative $Z$ direction. This justifies the role of the parameter $\alpha$ as the acceleration of the source particle(s) in the weak field regime.%\\ \\In the transformation \eqref{mink_transf} one has two choices of sign, $T = \pm\zeta \sinh \tau$ and $Z = \pm\zeta \cosh \tau $. Taking the negative transformation we get an origin particle of the accelerated frame which accelerates in the negative $Z$ direction with acceleration $\alpha$. This is because the C-metric actually describes \emph{two} black holes, accelerating away from each other. However, they are in causally different regions of spacetime (they are outside each others Rindler horizon), therefore we are free to choose one of them and focus on that region of spacetime which corresponds to choosing only one sign in the transformation \eqref{mink_transf}.

\section{A spherical accelerated shell}
\subsection{The spacetime inside and outside the shell} 
As we have seen, the C-metric can be used to describe an accelerating black hole or Schwarzschild-like particle. Our model is a shell which has this metric on the outside. This is a generalization of the Schwarzschild case, where one can have several sources for the exterior Schwarzschild solution, including a static shell.\\ \\
We seek an interior metric which is flat and able to produce perfect inertial dragging. This means that the inertial frames on the inside should accelerate with the same acceleration as the shell. %In the coordinates we use, $r=0$ is at all times the physical singularity associated with the position of the source of the C-metric. This means that these coordinates are accelerated relative to an unaccelerated observer.
We will therefore match the exterior coordinates to the interior coordinates such that the interior inertial frames follow the shell (these coordinates will be accelerated relative to some unaccelerated observer on the outside of the shell).\\ \\
It might be tempting to use the C-metric with $m=0$ inside the shell. As we have seen the metric then reduces to the metric of a uniformly accelerating reference frame. However, one needs to consider how the inertial frames inside the shell behave. Reference particles in a uniformly accelerating frame feel a uniform gravitational field. Hence they are not free. We are interested in a spacetime in which \emph{free particles accelerate along with the shell}. We must therefore choose Minkowski spacetime in spherical coordinates that follow the shell. Therefore the metric tensor inside the shell is:
\begin{align}
g_{\mu\nu} = \text{diag}\left( -1, 1, r^2, r^2\sin^2\theta \right)\:.
\label{minkowski}
\end{align}
We need the $\Gamma\indices{^r_{\mu\nu}}$ Christoffel symbols. The non-zero ones are:
\begin{align}
\Gamma\indices{^r_{\theta\theta}} = &  -r \\
\Gamma\indices{^r_{\phi\phi}} = & - r\sin^2\theta  \:.
\label{gammar_inside}
\end{align}
Outside the shell we have the C-metric given by \eqref{c_metric}. The $\Gamma\indices{^r_{\mu\nu}}$ Christoffel symbols we need are:
\begin{align}
 & \Gamma\indices{^r_{tt}}= Q\left[ \alpha^2(m-r) + \frac{m}{r^2} - \frac{\alpha \cos\theta Q}{\sqrt{D}} \right] \:, \\
 & \Gamma\indices{^r_{\theta\theta}}= -\frac{Qr}{P\sqrt{D}} \:, \\ 
 & \Gamma\indices{^r_{\phi\phi}}= -\frac{QPr\sin^2\theta}{\sqrt{D}(1+2\alpha m)^2} \:.
\end{align}

\subsection{Solving Lanczos' equation}
\subsubsection{The properties of the hypersurface}
We choose a hypersurface given by $r = R =$ constant. %This gives the restriction:
%\begin{align}
%\Phi(t,r,\theta,\phi) = r-R = 0 \:,\: R = \text{const.} 
%\end{align}
%The hypersurface is timelike since it consists of particles moving along timelike curves: $\epsilon = 1$. The derivative of the restriction is:
%\begin{align}
%\Phi_{,\alpha} = (0,1,0,0) \:.
%\end{align}
%This gives the normal vector:
%\begin{equation}
%n_\alpha = (0, n_r, 0,0)\:,
%\end{equation}
%with
%\begin{equation}
%n_r = \frac{1}{|g^{rr} \Phi_{,r}\Phi_{,r}|^{1/2}} = 
%\begin{cases}
%\hfill 1 \hfill, & r<R\:, \\
%\hfill \frac{1}{\sqrt{DQ}}, & r>R\:,
%\end{cases}
%\end{equation}
%and
%\begin{equation}
%n^r =g^{rr}n_r= 
%\begin{cases}
%\hfill 1 \hfill, & r<R\:, \\
%\hfill \sqrt{DQ}, & r>R\:.
%\end{cases}
%\end{equation}
Let $(\tau, \vartheta, \varphi)$ be the intrinsic coordinates of the hypersurface, let $(t^+, r^+, \theta^+, \phi^+)$ be the coordinates outside the surface and $(t^-, r^-,$ $\theta^-, \phi^-)$ be the coordinates inside. Then the surface is given by:
\begin{align}
t^- = \tau\:,\: &\theta^- = \vartheta\: ,\: \phi^- = \varphi \:, \\
r^- = & \,R = \text{ const.}
\end{align}
The tangent vectors in $\Sigma$ are:
\begin{align}
e^{\alpha^-}_\tau = (1,0,0,0) \:,\\
e^{\alpha^-}_\vartheta = (0,0,1,0) \:,\\
e^{\alpha^-}_\varphi = (0,0,0,1) \:.
\end{align}
The induced metric has % is:
%\begin{align}
%h_{ab} = g_{\alpha\beta}^\pm e^{\alpha^\pm}_a e^{\beta^\pm}_b\:.
%\end{align}
%With $g_{\mu\nu}^-$ (and $r^- = R$) we get 
the following non-zero components:
\begin{align}
h_{\tau\tau} & = -1\:, \notag \\
h_{\vartheta\vartheta} & = R^2\:, \label{induced_shell_metric} \\
h_{\varphi\varphi} & = R^2\sin^2\vartheta \:. \notag
\end{align}
Using the continuity of the induced metric %:
%\begin{equation}
%\left[ h_{ab}\right] = 0 \:,
%\label{continous_induced_metric}
%\end{equation}
we find the $e^{\alpha^+}_a$ tangent vectors:
\begin{align}
g_{\alpha\beta}^+ e^{\alpha^+}_a e^{\beta^+}_b = g_{\alpha\beta}^- e^{\alpha^-}_a e^{\beta^-}_b \:.
\end{align}
This gives the following tangent vectors (choosing the positive solution):
\begin{align}
e^{\alpha^+}_\tau &= \left(\sqrt{\frac{D}{Q}},0,0,0\right) \:,\\
e^{\alpha^+}_\vartheta &= \left(0,0,\sqrt{PD},0\right) \:,\\
e^{\alpha^+}_\varphi &= \left(0,0,0,\frac{\sin\vartheta (1+2\alpha m)}{\sin\theta^+}\sqrt{\frac{D}{P}}\right) \:.
\end{align}
The relationship between the exterior and intrinsic coordinates are:
\begin{align}
\tau &= \sqrt{\frac{Q}{D}} t^+\:,\: d\vartheta = \frac{1}{\sqrt{PD}}\, d\theta^+\:,\: \varphi = \frac{\sin\theta^+}{\sin\vartheta (1+2\alpha m)}\sqrt{\frac{P}{D}}\,\phi^+ \:.
\end{align}
The integration over $\theta^+$ involves an elliptical integral, the relationship between $\vartheta$ and $\theta^+$ is therefore expressed in differential form.

\subsubsection{The energy-momentum tensor of the shell}
Using equations \eqref{excurve} and \eqref{intr_transf} yields the following components of the extrinsic curvature tensor in the intrinsic coordinates:
%
%In $\mathcal{M}^-$ the relevant components of the extrinsic curvature tensor in the interior coordinate system read:
%\begin{align}
%K_{tt^-}^- %&= n_r^-\Gamma\indices{^r_{tt}^-} \notag \\
%& = 0 \:,\\
%K_{\theta\theta^-}^-% &=  n_r^-\Gamma\indices{^r_{\phi\phi}^-} \notag \\
%&= -r\:, \\
%K_{\phi\phi^-}^- %&=  n_r^-\Gamma\indices{^r_{\theta\theta}^-}\notag \\
%&= -r\sin^2\theta\:.
%\end{align}
%The extrinsic curvature tensor in $\mathcal{M}^+$, in the exterior coordinate system, has the following relevant components:
%\begin{align}
%K_{tt^+}^+ %&= n_r^+\Gamma\indices{^r_{tt}^+} \notag \\
%& = \sqrt{\frac {Q}{D}}\left[
%\alpha^2 (m-r) + \frac{m}{r^2} - \frac{\alpha\cos\theta^+ Q}{\sqrt{D}}
%\right] \:,\\
%K_{\theta\theta^+}^+ %&=  n_r^+\Gamma\indices{^r_{\phi\phi}^+} \notag \\
%&= -\frac{\sqrt{Q}r}{PD}\:, \\
%K_{\phi\phi^+}^+% &=  n_r^+\Gamma\indices{^r_{\theta\theta}^+} \notag \\
%&= -\frac{\sqrt{Q}Pr^2 \sin^2\theta^+}{D(1+2\alpha m)^2}\:.
%\end{align}
%This transforms according to eq. \eqref{int_excurve} to the intrinsic coordinates:
%\begin{align}
%K_{\tau\tau}^- = 0 \:,\: K_{\vartheta\vartheta}^- = -R \:,\: K_{\varphi\varphi}^- = -R\sin^2\vartheta \:,
%\end{align}and
%\begin{align}
%K_{\tau\tau}^+ &= \sqrt{\frac {D}{Q}}\left[
%\alpha^2 (m-R) + \frac{m}{R^2} - \frac{\alpha\cos\theta^+ Q}{\sqrt{D}}
%\right] \:, \\
%K_{\vartheta\vartheta}^+ &= -R\sqrt{Q} \:, \\
%K_{\varphi\varphi}^+ &= -R\sin^2\vartheta \sqrt{Q} \:.
%\end{align}
%Raising the indices gives the mixed components:
\begin{align}
%K\indices{^\tau_\tau^-} = 0\\
%K\indices{^{\tau^-}_\tau} = 0\\
{K\indices{^\tau}}^-_\tau =  0 \:,\: {K\indices{^\vartheta}}_\vartheta^- &= -\frac{1}{R} = {K\indices{^\varphi}}_\varphi^- \:,
\end{align}and
\begin{align}
{K\indices{^\tau}}_\tau^+ &= - \sqrt{\frac {D}{Q}}\left[
\alpha^2 (m-R) + \frac{m}{R^2} - \frac{\alpha\cos\theta^+ Q}{\sqrt{D}}
\right] \:, \\
{K\indices{^\vartheta}}_\vartheta^+ &=-\frac{\sqrt{Q}}{R} \\
 &= {K\indices{^\varphi}}_\varphi^+ \:.
\end{align}
The discontinuities of the extrinsic curvature tensor are:
\begin{align}
\Bigl[ K\indices{^\vartheta_\vartheta}\Bigr] %& = {K\indices{^\vartheta}}_\vartheta^+ - {K\indices{^\vartheta}}_\vartheta^-  \notag \\
& = \frac{1}{R}\left(1-\sqrt{Q}\right) \\
& = \Bigl[ K\indices{^\varphi_\varphi} \Bigr] \:,\\%\end{align}and\begin{align}
\Bigl[ K\indices{^\tau_\tau}\Bigr]% & ={K\indices{^\tau}}_\tau^+ - {K\indices{^\tau}}_\tau^-  \notag \\
& = -\sqrt{\frac{D}{Q}}\left[ 
\alpha^2(m-R) +\frac{m}{R^2} - \frac{\alpha\cos\theta^+ Q}{\sqrt{D}} \right]\:.
\end{align}
From the Lanczos equation \eqref{lanczos} we have:
\begin{align}
S\indices{^\tau_\tau} %& = \frac{1}{8\pi} \left( \Bigl[ K\indices{^\tau_\tau} \Bigr] - \delta\indices{^\tau_\tau} \Bigl[ K \Bigr]\right) \notag  \\
& = \frac{1}{8\pi} \left( 
 - \Bigl[ K\indices{^\vartheta_\vartheta} \Bigr]-\Bigl[K\indices{^\varphi_\varphi} \Bigr]\right) \:,\\
S\indices{^\vartheta_\vartheta} & = \frac{1}{8\pi} \left( 
  \Bigl[ K\indices{^\vartheta_\vartheta} \Bigr]-\Bigl[K\Bigr]\right)\notag  \\
%& =  \left(  - \Bigl[ K\indices{^\tau_\tau} \Bigr]-\Bigl[K\indices{^\varphi_\varphi} \Bigr]\right)  \\
%& =  \left(  - \Bigl[ K\indices{^\tau_\tau} \Bigr]-\Bigl[K\indices{^\vartheta_\vartheta} \Bigr]\right)  \notag \\
& = S\indices{^\varphi_\varphi} \:,
\end{align}
with the result:
\begin{subequations}
\begin{empheq}[box=\widefbox]{align}
 S\indices{^\tau_\tau} &  = -\frac{1}{4\pi R} \left(1-\sqrt{Q}\right)  \\
S\indices{^\vartheta_\vartheta} & = \frac{1}{8\pi } \left(
\frac{-1+\sqrt{Q}}{R} \!+ \!\sqrt{\frac{D}{Q}}\!\left(
\!\alpha^2(m\!-\!R) \!+\!\frac{m}{R^2} \!-\! \frac{\alpha\cos\theta^+ Q}{\sqrt{D}}
\right)\!\right) \\ 
& = S\indices{^\varphi_\varphi} \:.
\end{empheq}
\label{endeligsvar}
\end{subequations}

\subsection{Physical interpretation of the shell}
The energy-momentum tensor of a perfect fluid can be written:
\begin{equation}
S^{ab} = (\sigma + p) u^a u^b + p h^{ab}\:.
\end{equation}
Since the shell is comoving with the coordinates, the spatial velocity components are zero:
\begin{equation} 
u^a = (u^\tau, 0, 0)\:.
\end{equation}
This gives
\begin{align}
S\indices{^\tau_\tau} & = -\sigma \:,\\
S\indices{^\vartheta_\vartheta} & = p \\
&= S\indices{^\varphi_\varphi} \:.
\end{align}
Comparing this to our final expression for $S\indices{^a_b}$ in \eqref{endeligsvar} leads to:
\begin{align}
\sigma & = \frac{1}{4\pi R}\left( 1-\sqrt{Q} \right)\:, \label{c_source1}\\
p  & = \frac{1}{8\pi R}\left(
-\left(1-\sqrt{Q}\right) + R\sqrt{\frac{D}{Q}}\left(
\alpha^2(m-R) + \frac{m}{R^2} - \frac{\alpha \cos\theta^+ Q}{\sqrt{D}}
\right)\right) \label{c_source2}\,.
\end{align}
From this we can conclude that the shell consists of a perfect fluid with proper rest mass density $\sigma$ and pressure $p$. In the limit $R\rightarrow 2m$, the function $Q$ goes to zero and we see that the rest mass density approaches the finite value of $(4\pi R)^{-1} = (8\pi m)^{-1}$, while the second term in the pressure diverges. This divergence is not a coordinate effect, but rather a manifestation of the fact that infinite pressure is needed in order to keep a shell at rest when it is located exactly at the Schwarzschild horizon. At this distance, space itself (or the ``river of space'' \cite{rivermodel}) flows inward towards the center of the black hole with the speed of light.\\ \\The pressure contributes to the total gravitational mass. However, integrating the relativistic mass density is problematic since the pressure is a function of the exterior polar angle. Just from the fact that we know that the pressure contributes, and that the pressure diverges as $R\rightarrow 2m$ we can conclude that just as in the Schwarzschild case, we cannot expect a perfect fluid to remain static in this limit. \\ \\%The equations \eqref{sigma_eq} and \eqref{constant_sigma_p} are still valid since only a perfect fluid assumption and the induced metric (which is the same as in the Schwarzschild case) was used in their derivation. However the constant on the right hand side in \eqref{constant_sigma_p} is now only a constant with respect to time, since the pressure varies with the polar angle. This means that in the static case we still have conservation of rest mass, and in the general case we still have that $R^2(\sigma+p) = const.$ (in time).\\ \\
The accelerated shell with mass density and pressure given by equations \eqref{c_source1} and \eqref{c_source2} is a new source of the C-metric. By construction there is perfect inertial dragging inside this shell. \\ \\ \textbf{The limit $m\rightarrow 0$:} \\When the parameter $m$ is set to zero the resulting rest mass density and pressure are:
\begin{align}
\sigma & = \frac{1}{4\pi R}\left( 1-\sqrt{1-\alpha^2 R^2} \right)\:, \\
p  & = \frac{1}{8\pi R}\left(
-\left(1-\sqrt{1-\alpha^2 R^2}\right) + \frac{-\alpha R(\alpha R + \cos\theta^+)}{\sqrt{1-\alpha^2 R^2}}
\right) \:.\label{massless_pressure}
\end{align}
These expressions are not vanishing. This shows that there still is a massive shell present, it is only the mass parameter $m$ which is zero. \\ \\The Tolman-Whittaker mass \cite{gronhervik, tolman} is non-zero:
\begin{align}
M &=  \int_V \left(T\indices{^{\alpha}_{\alpha}} - 2 T\indices{^{0}_{0}} \right) \sqrt{-g} dV \\
&= \int_V \left(-T\indices{^{t}_{t}} + T\indices{^{r}_{r}}  +T\indices{^{\theta}_{\theta}} + T\indices{^{\phi}_{\phi}} \right) \sqrt{-g} dV \:.\label{tolman_whittaker} \\
&= \int_0^R \int_0^\pi \int_0^{2\pi} \delta(r-R) \left(\sigma + 2 \frac{r^2}{R^2} p(\theta^+) \right)r^2 \sin\theta dr d\theta d\phi\label{TW_massless_limit}
\end{align}
The first term in \eqref{massless_pressure} will cancel the $\sigma$ term in the integral, and we are left with integration over the last term in \eqref{massless_pressure}. The integration over the $\theta^+$ dependent term cannot be done analytically since $\theta^+$ is only given as an implicit funtion of $\theta$ which cannot be inverted. A numerical integration shows that $M>0$ \cite{ba}. \\ \\
This means that the parameter $m$ only represents the gravitational mass in the limit $\alpha\rightarrow 0$. In the general case a non-zero $\alpha$ affects the gravitational mass. The exact form of this dependence remain unknown since performing the Tolman-Whittaker integration cannot be done analytically in the general case.\\ \\\textbf{The limit $\alpha \rightarrow 0$:}\\In this limit we recover the unaccelerated Schwarzschild shell, and the mass parameter $m$ is then equal to the gravitational mass.

\section{Discussion}
\begin{figure}[h]
\centering
\includegraphics[width=0.9\textwidth]{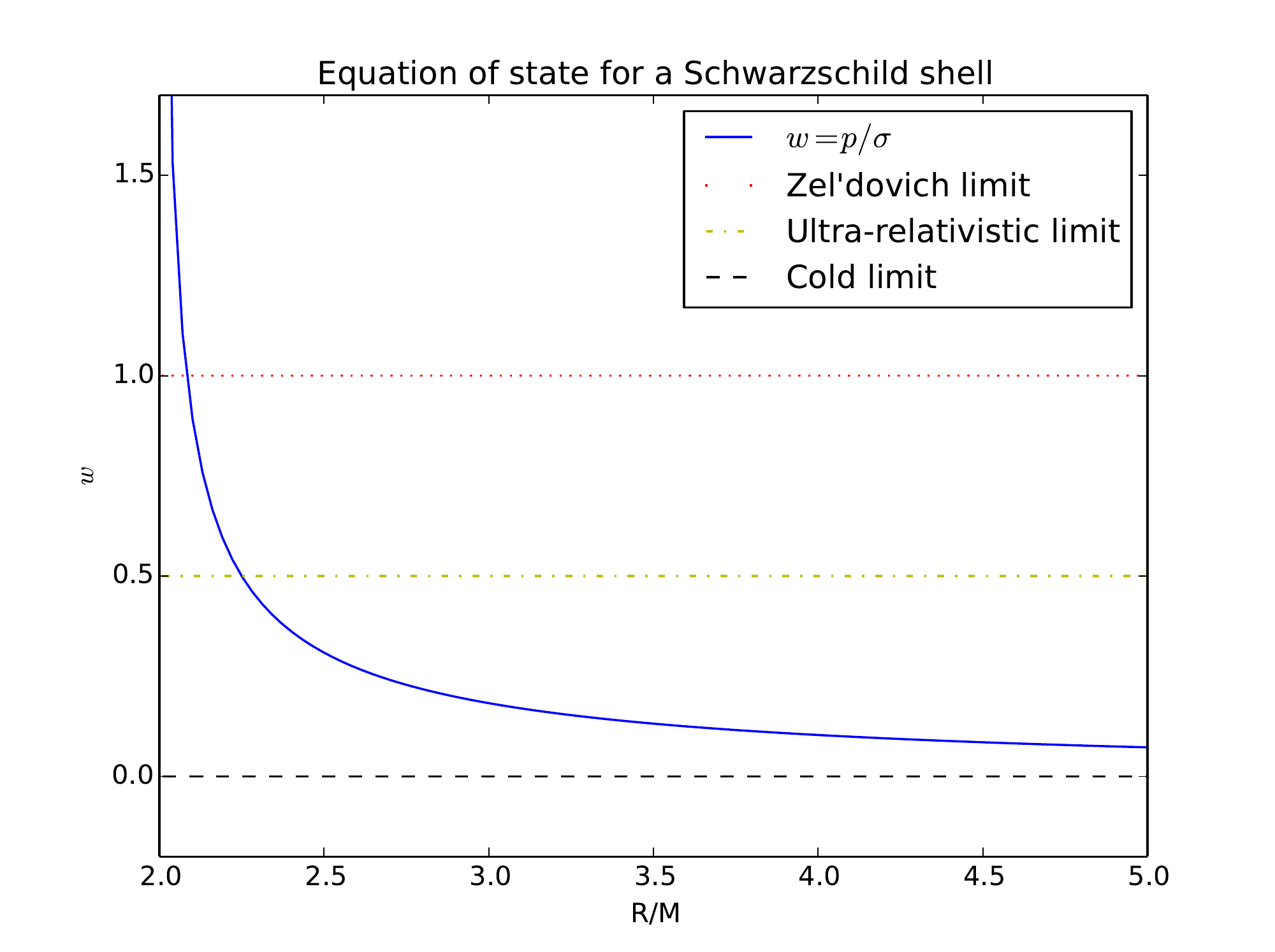}
\caption{The equation of state for a perfect fluid Schwarzschild shell.}
\label{fig: eq_of_state_sch}
\end{figure}
\begin{figure}[h]
\centering
\includegraphics[width=0.9\textwidth]{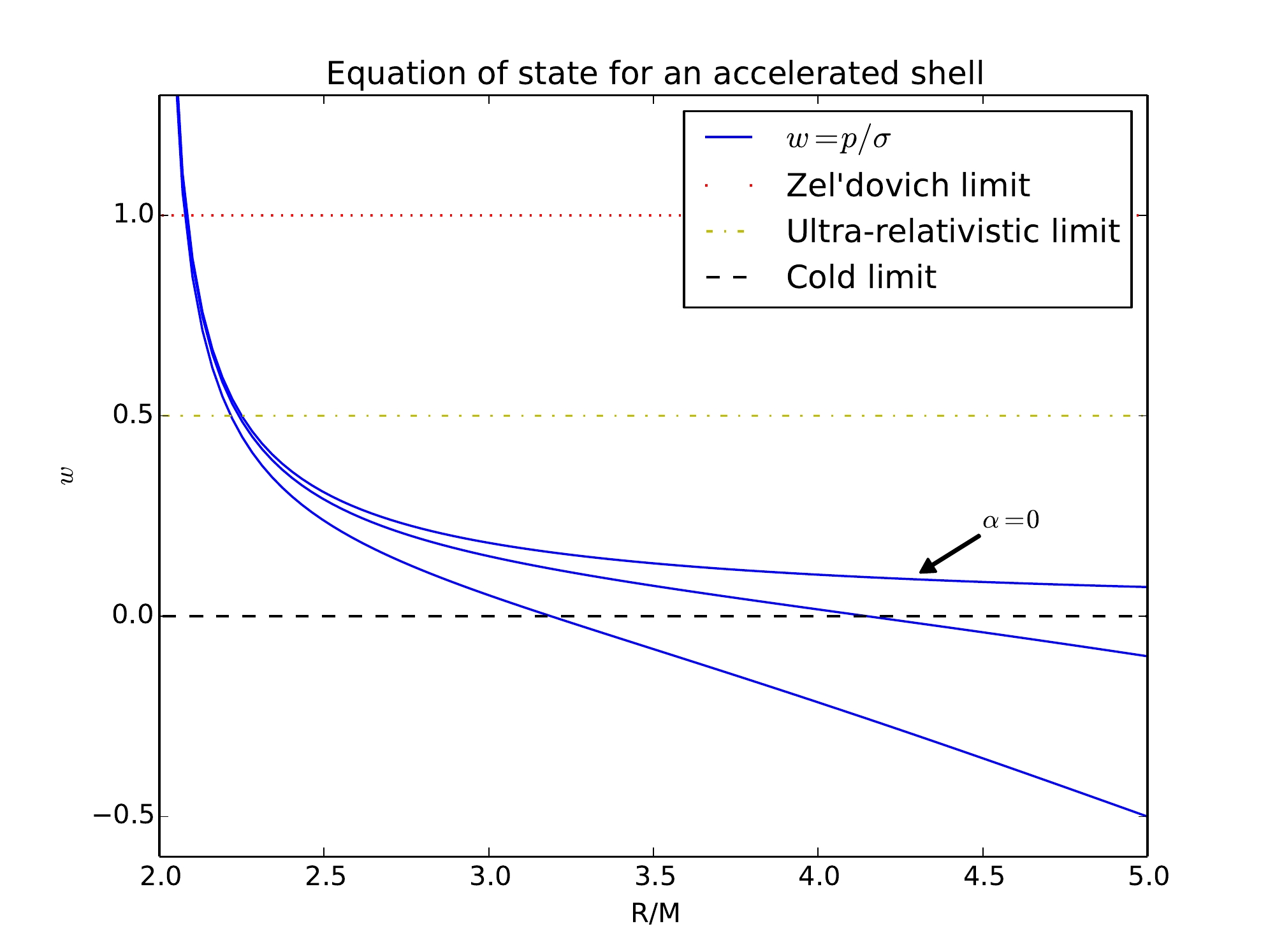}
\caption{The equation of state for the accelerated shell along $\cos\theta^+ = 0$. $\alpha=0$ corresponds to the Schwarzschild case of figure \ref{fig: eq_of_state_sch}. The two lower blue curves represent increasing $\alpha$, 0.05 and 0.1 respectively.}
\label{fig: eq_of_state_acc_shell}
\end{figure}
\begin{figure}[h]
\centering
\includegraphics[width=0.9\textwidth]{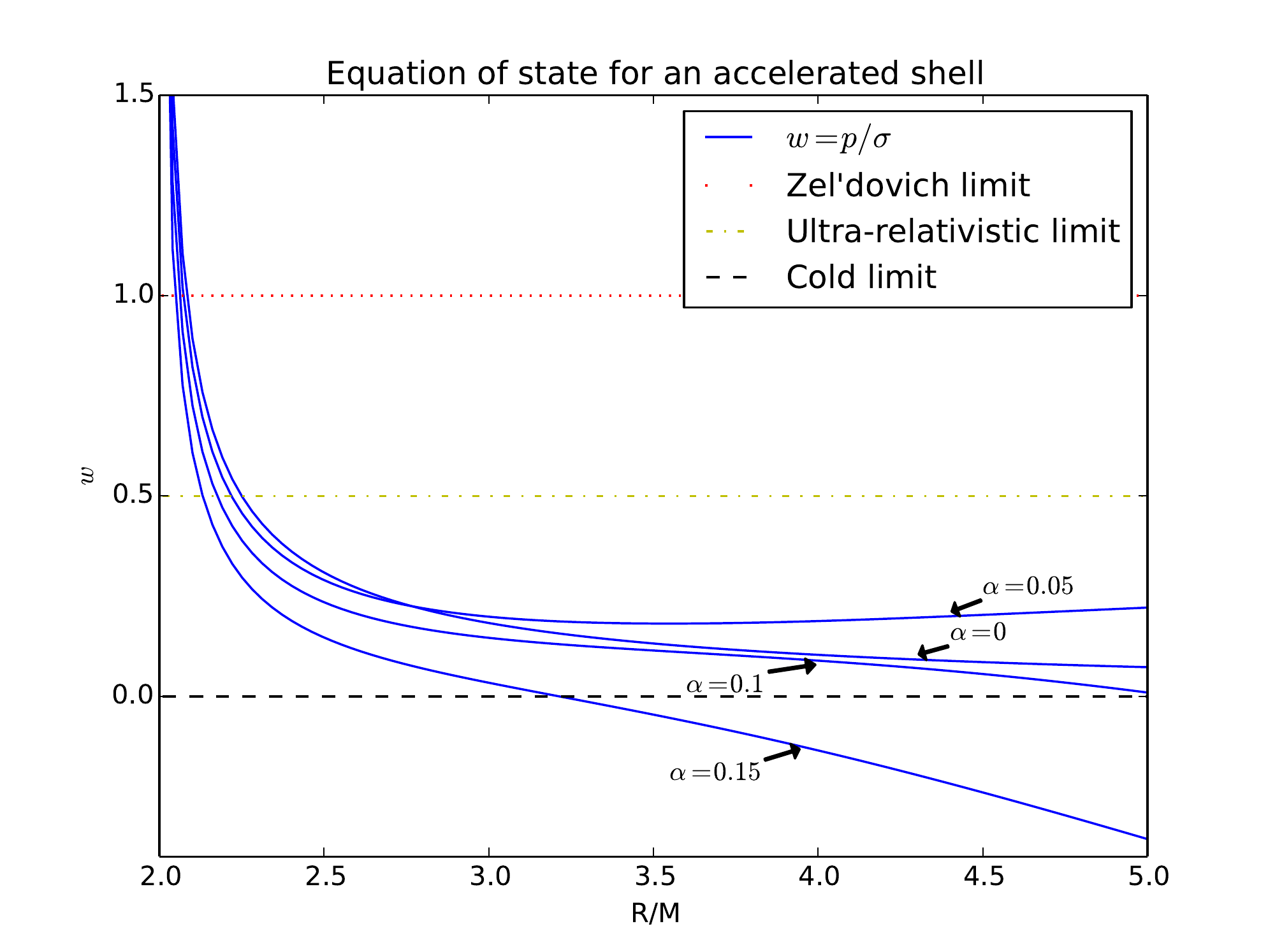}
\caption{The equation of state for the accelerated shell along $\cos\theta^+ = -1$. $\alpha=0$ corresponds to the Schwarzschild case of figure \ref{fig: eq_of_state_sch}. The other blue curves represents $\alpha= 0.05$ , 0.1 and 0.15.}
\label{fig: eq_of_state_acc_shell_cosminone}
\end{figure}
\begin{figure}[h]
\centering
\includegraphics[width=0.9\textwidth]{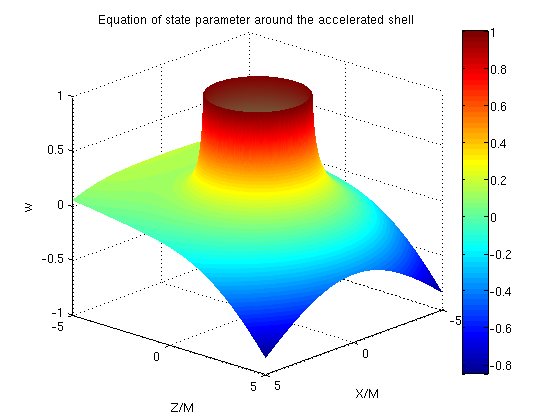}
\caption{The equation of state parameter $w$ along varying $\theta^+$ for an accelerated shell with a radius $R$. The shell accelerates in the negative $Z$-direction ($Z=R\cos\theta^+$). $\alpha = 0.05$ is used. The pressure is constant along the $\phi^+$ direction. There is a cut-off near the horizon where the pressure diverges.}
\label{fig: eq_of_state_surfaceplot}
\end{figure}
\begin{figure}[h]
\centering
\includegraphics[width=0.9\textwidth]{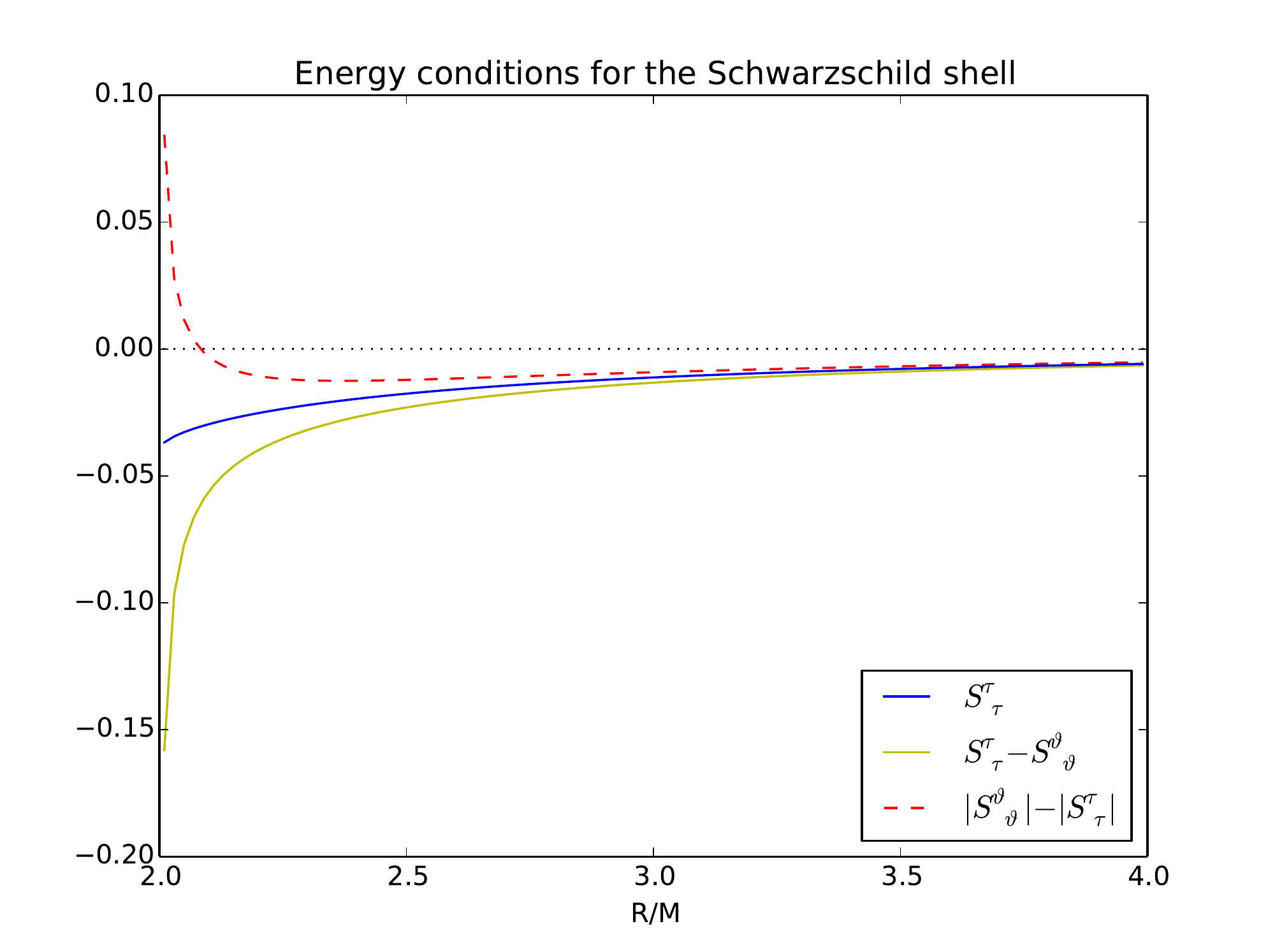}
\caption{The two solid lines represent the two functions that must be below zero for the weak energy condition to hold. The red dashed line represents the function that must be below zero for the dominant energy conditions to hold.}
\label{fig: energy_conditions_sch}
\end{figure}
\begin{figure}[h]
\centering
\includegraphics[width=0.9\textwidth]{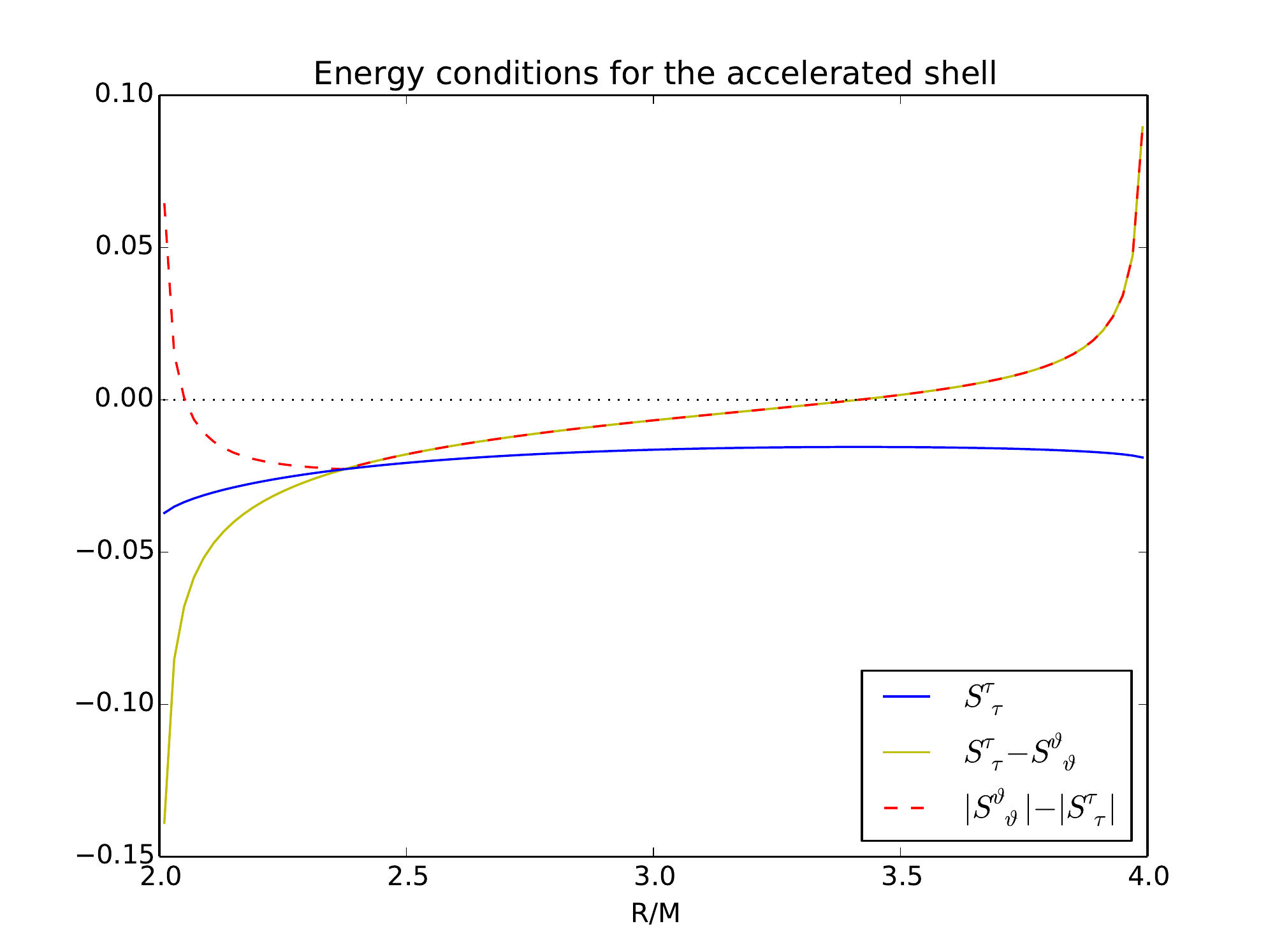}
\caption{The two solid lines represent the two functions that must be below zero for the weak energy condition to hold. The red dashed line represents the function that must be below zero for the dominant energy conditions to hold. For the angle-dependent $S\indices{^\vartheta_\vartheta}$ we again chose the ``equatorial'' plane $\cos\theta^+ = 0$. Qualitatively the results are the same for all angles. $\alpha=0.25$ is chosen so $R/M=4$ corresponds to the acceleration horizon.}
\label{fig: energy_conditions}
\end{figure}
\subsection{The state of the shell}
In figure \ref{fig: eq_of_state_sch} we see that the static Schwarzschild shell can be desribed as a perfect fluid with a non-zero positive pressure. Radiation, or a gas consisting of ultra-relativistic particles (particles where the energy from the rest mass is negligible compared to the kinetic/total energy of the particles), have a traceless energy-momentum tensor, $T\indices{^\alpha_\alpha} = 0$. Since $T\indices{^r_r} = 0$ an ultra-relativistic shell has equation of state
\begin{align}
p = \frac{1}{2} \sigma \:.
\end{align}
We see that when $R = \frac{9}{4}M$, the pressure is so large that the particles of the shell become ultra-relativistic. Past this limit the shell can still be kept static with a finite pressure. An absolute upper limit for the parameter $w$ in general relativity is given by Zel'dovich \cite{zel}. A physical limit on the speed of sound in a material is the speed of light. In a material where these are equal, $w$ is equal to one. According to Zel'dovich this is the upper limit alowed by general relativity. For the Schwarzschild shell this accours at $R = \frac{25}{12}M$. Past this a static schell cannot be interpreted within the general theory of relativity. In the collapse-limit the pressure diverges and one cannot expect the shell to remain static.\\ \\In figure \ref{fig: eq_of_state_acc_shell} we see the equation of state for the accelerated shell. Since the pressure is a function of the polar angle the ``equatorial'' plane $\cos\theta^+ = 0$ is chosen for convenience. Increasing the acceleration parameter $\alpha$ lowers the pressure along this curve so much that the material obtains negative pressures for larger radia. This can be interpreted as being due to the acceleration from the source. The source of the acceleration is a cosmic string, and like the LIVE (Lorentz Invariant Vacuum Energy) represented by the cosmological constant, this can cause repulsive gravity. At the zero-points on the graph this repulsive tendency cancels the ordinary pressure in the material, making the shell effectivey a shell of dust. However, the pressure is only zero along $\cos\theta^+ = 0$, so the whole shell will not behave as dust.\\ \\We also see that the zero-point of the pressure increases as $\alpha$ decreases, and will go to infinity as $\alpha$ goes to zero. This means that a static shell of dust is only realized in the limit where both $\alpha\rightarrow 0$ and $R\rightarrow\infty$. \\ \\In figure \ref{fig: eq_of_state_acc_shell_cosminone} we see the same plot of the equation of state for the shell, now for $\cos\theta^+ = -1$, with one added value of $\alpha = 0.15$. At this angle the pressure is positive for $\alpha = (0, 0.05, 0.1)$, only $\alpha = 0.15$ becomes negative inside the domain. At larger radia the pressure increases for small $\alpha$ and decreases again for larger $\alpha$. A more complete picture of the equation of state parameter $w$ is shown in figure \ref{fig: eq_of_state_surfaceplot}. The previous plots corresponds to vertical slices through the axes of this surface plot. $\cos\theta^+ = 0$ corresponds to a slice perpendicular to the $Z$-axis and $\cos\theta^+ = \pm1$ corresponds to a slice perpendicular to the $X$-axis. \\ \\The fact that the coefficient $w$ is not a constant but depends on the polar angle questions the validity of the perfect fluid assumption. It remains an open question if it is possible to choose a non-spherical shape for the shell which removes this feature. \\ \\In figure \ref{fig: energy_conditions_sch} and \ref{fig: energy_conditions} we see plots of the energy condition functions - the functions that need to be below zero for the relevant energy conditions to hold \cite{ba, pfister}. The two solid curves represent the function that need to be negative for the weak energy condition to hold. The dashed red curve represents the function that must be negative for the dominant energy condition to hold. We see that the weak energy condition is obeyed in the interesting regions near the collapse limit. The dominant energy condition is violated near the collapse limit. In fact, the violation occours exactly at the Zel'dovich limit, since the dominant energy condition for our shell reads
\begin{align}
\left| S\indices{^\vartheta_\vartheta} \right| - &\left| S\indices{^\tau_\tau} \right|  \leq 0 \:, \notag \\
\left|\, p \,\right| - &\left|\, \sigma \,\right| \,\leq 0 \:.
\end{align}
The highest allowed value for the pressure according to this is $p=\sigma$, which is the Zel'dovich limit.\\ \\
The fact that the energy conditions are violated near the acceleration horizon for the accelerated shell does not appear to be so problematic since it is in the other limit that we expect our model to give rise to perfect inertial dragging. Comparing these plots with figure 1 in \cite{pfister} it seems that the violation of the dominant energy condition near the collapse limit might be overcome by giving the shell an electric charge (such a solution exists and is referred to as the charged C-metric). Wether this is true however remains an open question.

\subsection{Inertial dragging}
%Is the fact that we have Minkowski spacetime inside the shell enough to ensure that perfect inertial dragging is realized? Or should one make some coordinate transformation to an unaccelerated observer somewhere outside the shell, and calculate the observed acceleration of a test particle inside the shell? (Expecting the result to be 1. It might depend on $M/R$. Ideally it would go to 1 as $R\rightarrow 2M$.)\\ \\Can we make the argument that this is actually what you will get? \\ \\
The coordinates inside the shell coincide with those on the shell. Hence the interior coordinates are comoving with the shell. Since there is Minkowski spacetime inside the shell, a particle at rest in the interior coordinate system is a free particle. As viewed by a non-accelerated observer relative to an observer at rest in the asymptotic region far away from the shell, such a particle accelerates together with the shell. This means that there is perfect translational inertial dragging inside the shell.
\\ \\Our model has one strange property. There is perfect inertial dragging inside the shell independent of the value of $m/R$. This is not realistic. One cannot expect that there is perfect inertial dragging inside a shell with a large radius and a very small mass. \\ \\%This means that we have described a shell where perfect inertial dragging is realized, but we have not deduced that perfect inertial dragging is the result of the presence of such an accelerated mass shell.
The shell with energy-momentum tensor given by eq. \eqref{endeligsvar} is constructed to produce perfect inertial dragging for all allowed values of the mass parameter $m$. This is not necessarily true for a mass shell accelerated by other means with a different energy-momentum tensor. By analogy to the rotational case, we expect that an arbitrary accelerated shell produces perfect inertial dragging in the limit $R\rightarrow 2m$. \\ \\As mentioned earlier, giving the shell a charge might salvage the violated dominant energy condition. However, from studying figure 4 in \cite{pfister}, we see that in the part of parameter space obeying the dominant energy condition, perfect inertial dragging is not realized. 

\section{Conclusion and outlooks}
In the present paper the physical properties of an accelerated static mass shell as a source of the (external) C-metric has been described. By construction, perfect inertial dragging is realized inside the shell. In this way our model casts light on the requirements for achieving perfect translational inertial dragging. In addition, it provides a source for the C-metric. Our results are analytical and exact to all orders of the acceleration parameter $\alpha$. \\ \\
The physical properties of the shell have been analyzed and we found that the weak energy condition holds in the interesting regions (when the radius of the shell is much less than the acceleration horizon). However, near the collapse limit, the dominant energy condition is violated. This might be overcome by giving the shell a charge. To do this one needs to extend our external metric to the charged C-metric. However, in \cite{pfister}, this invalidated their perfect inertial dragging condition.\\ \\
The equation of state parameter $w$ depends on $\theta^+$. This makes a perfect fluid assumption questionable. This might possibly be repaired by introducing a non-spherical shape for the shell. \\ \\
Near the collapse limit, the pressure diverges. This comes from the fact that we have required a static mass shell. At the Schwarzschild horizon we cannot expect a mass shell to remain static. This is analogous to the rotational case. \\ \\
The main point is that a shell of matter is only an idealized model for a Universe with lookback distance equal to the Schwarzschild radius of the mass inside this distance \cite{brillcohen, grontwin}. One does not need this static model to behave perfectly well in this limit since this model is not realized in nature. However when such models give rise to perfect inertial dragging, it becomes very plausible that the Universe also gives rise to perfect inertial dragging - a condition necessary for the fulfilment of Mach's principle
\printbibliography

\end{document}